\documentclass[11pt]{article}
\usepackage{epsf,amsfonts}
\usepackage{fontenc,indentfirst, delarray,amsfonts,amsmath,amssymb}
\usepackage{graphicx}
\DeclareGraphicsExtensions{ps,eps,ps.gz} \headheight 1cm
\makeatletter
\def\@citex[#1]#2{\if@filesw\immediate\write\@auxout
        {\string\citation{#2}}\fi
\def\@citea{}\@cite{\@for\@citeb:=#2\do
        {\@citea\def\@citea{,}\@ifundefined
        {b@\@citeb}{{\bf ?}\@warning
        {Citation `\@citeb' on page \thepage \space undefined}}
        {\csname b@\@citeb\endcsname}}}{#1}}
\newif\if@cghi
\def\cite{\@cghitrue\@ifnextchar [{\@tempswatrue
        \@citex}{\@tempswafalse\@citex[]}}
\def\citelow{\@cghifalse\@ifnextchar [{\@tempswatrue
        \@citex}{\@tempswafalse\@citex[]}}
\def\@cite#1#2{{\if@cghi\unskip$\null^{#1}$\else #1\fi\if@tempswa\typeout
        {warning: optional citation argument ignored: `#2'} \fi}}

\def\@biblabel#1{$\null^{#1}$}
\makeatother

\newcommand{\norm}[1]{\left\lVert#1\right\rVert}
\newcommand{\abs}[1]{\lvert#1\rvert}

\newcommand{\scl}[2]{\langle#1,#2\rangle}

\def\dm{\lp\begin{array}}
\def\fm{\end{array}\rp}
\def\dbb{\lb\begin{array}}
\def\fbb{\end{array}\rb}
\def\dbn{\left.\begin{array}}
\def\fbn{\end{array}\right.}

\def\ee{{\cal E}}
\def\lb{\left[}
\def\rb{\right]}
\def\lp{\left(}
\def\rp{\right)}

\def\m3{M_3 \lp \cc \rp}
\def\m2{M_2 \lp \cc \rp}

\def\cc{{\mathbb{C}}}
\def\rr{{\mathbb{R}}}

\def\ii{{\mathbb{I}}}

\def\aa{{\cal A}}

\def\dd{{\cal D}}
\def\hh{{\cal H}}

\def\oo{{\cal O}}
\def\ss{{\cal S}}
\def\mm{{M}}

\def\hhh{{\mathbb H}}

\def\cinf{C^{\infty}\lp\mm\rp}

\def\L2{L_2(\mm)}

\def\ot{\otimes}

\def\xo0{\omega^0_x}
\def\yo0{\omega^0_y}

\def\o0{\omega_0}

\def\xo0{x_\omega^0}
\def\yo0{y_\omega^0}

\begin{document}


%
%

\title{ What kind of noncommutative geometry for quantum gravity
?}
\date{ }
\author{{\small Pierre Martinetti}
\\
\\
{\small Dpt. de Matem\'atica, Instituto Superior T\'ecnico, Lisboa, 1049-001, Portugal}\\
{\small Centre de physique th\'eorique, CNRS Luminy, 13288
Marseille France}\\
{\small {\em pmartin@math.ist.utl.pt}}}  \maketitle

\begin{abstract}
 We give a brief account of the description of the standard model in
noncommutative geometry as well as the thermal time hypothesis,
questioning their relevance for quantum gravity.
\end{abstract}


\section{Introduction}

First of all let us emphasize the importance of the question mark
in the title. Our aim is of course not to answer the question but
rather to underline that depending on the community one is talking
to, "noncommutative geometry" does not have the same meaning. In
most of the applications related to quantum gravity (whatever
candidate for a quantum theory of gravity is taken into account)
noncommutative geometry is understood as the geometry of a quantum
space, that is to say a space (spacetime or a phase space) whose
coordinates do not commute, either in a canonical way
\begin{equation}
[x^\mu, x^\nu] = \theta_{\mu\nu}
\label{canonical}
\end{equation}
where $\theta_{\mu\nu}$ is a constant, or in a Lie-algebraic way
\begin{equation}
\label{noncanonical} [x^\mu, x^\nu] = C^{\mu\nu}_\beta x^\beta.
\end{equation}
However the physical and/or geometrical meaning of those quantum
spaces is not entirely clear. For instance, passing into the
noncommutative realm, what happens to the most intuitive notions
in geometry such as points, distances, or to more elaborated
geometrical tools such as differential structure, homology or
spin?

From a more fundamental point of view Noncommutative
geometry\cite{connes} is an extension of geometry beyond the scope
of riemannian spin manifold. The later is encompassed as a
particular case, commutative, of the general theory. The price for
this generalization is a more abstract approach to geometry in
terms of spectral datas. Connes has first observed that the
geometrical information of a riemannian spin manifold can be
recovered from algebraic datas, the so called {\it spectral
triple} consisting in an algebra $\aa$, an Hilbert space $\hh$ and
an operator $D$ satisfying precise conditions\cite{gravity}. The
involved algebra is the algebra of smooth functions on the
manifold, which is commutative. Conversely a spectral triple with
$\aa$ commutative is associated to a spin manifold,
\begin{equation}
\text{Riemannian spin geometry} \Longleftrightarrow
\text{Commutative algebra}.
\end{equation}
Now the tools allowing to go from the right side of the arrow
(algebra) to the left side (geometry) do not rely on the
commutativity of the algebra. They are still available when the
algebra is not commutative. Hence a {\it noncommutative geometry}
is the mathematical object that one obtains starting from a
spectral triple $(\aa,\hh, D)$ in which the algebra $\aa$ is non
necessarily commutative,
\begin{equation}
\label{shortarrow}
 \text{Noncommutative geometry} \Longleftarrow
\text{Noncommutative algebra}
\end{equation}
It is not obvious that all quantum spaces considered in physics
literature can be described in spectral terms. However one can
expect that the further the physics of those spaces is
investigated, the deeper should be the mathematical coherence of
the underlying geometry. Hence it is likely that the physics of
quantum spaces be confronted sooner or later to some mathematical
questions addressed by Connes theory. For instance, as far as I
know, the notion of distance is not always available in quantum
spaces (a two-form metric is available, but its interpretation as
a line element and integration as a distance is not always
possible\cite{madore}). Nevertheless there are now more and more
point of contacts between various approaches to noncommutativity.
Recently several spectral triples have been proposed for quantum
spaces well known in deformation quantization (the Moyal
plane\cite{moyal}) or in quantum groups (the Podles
sphere\cite{podles}).  Our point here is not to be exhaustive but
only to give a brief account of Connes' theory as well as some of
its interest for physics. We shall focus on the description of the
standard model of elementary particles on the one hand, and on the
beautiful idea (related to the issue of time in relativity) that
"Von Neumann algebras naturally evolve with time" on the other
hand. As a conclusion we go back to the title of the talk and
question the interface of noncommutative geometry with quantum
gravity.

\section{Tools}

\subsection{Topology}

The degree zero of geometry is the ability to determine whether
two points are close to each other or not. This is the subject of
topology. A famous theorem by Gelfand, Naimark and Segal
establishes that the topological information of a compact space
$X$ is entirely contained within the algebra $C(X)$ of complex
continuous functions on $X$. As an algebra $C(X)$ is commutative
\begin{equation}
fg(x) \doteq f(x)g(x) = g(x)f(x) = gf(x),
\end{equation}
equipped with an involution *
\begin{equation}
f^*(x)\doteq \overline{f(x)}
\end{equation}
and a norm
\begin{equation}
\norm{f} \doteq \underset{x\in X}{\sup} \abs{f(x)}
\end{equation}
which make of $C(X)$ a $C^*$-algebra (i.e. closed in the norm
topology and such that $\norm{f}^2 = \norm{f^*}\norm{f}$).
Conversely  given a commutative $C^*$-algebra (with unit) $\aa$ it
is always possible to build a compact space such that $\aa$
interprets as the algebra of continuous functions on $X$. Hence
\begin{equation}
\label{cat}
\begin{array}{c}
\text{Commutative $C^*$-algebra with unit}\\
\text{$\aa$}\end{array} \Longleftrightarrow \begin{array}{c}
\text{Compact topological space}\\
\text{$X$}\end{array}
\end{equation}
Strictly speaking the equation above is an equivalence of
categories. For our purpose it is enough to understand how one
goes from one side to the other. We already know half of the
bridge
\begin{equation}
\label{bridge1}
 C(X) \leftarrow X.
\end{equation}
The other half is built on {\it characters} of the algebra, that
is to say homorphisms $\mu$ from $\aa$ to $\cc$
\begin{equation}
\mu(ab) = \mu(a)\mu(b) \quad \forall\, a,b\in\aa.
\end{equation}
The set $K(\aa)$ of characters of a commutative $C^*$-algebra with
unit is a compact space, hence the other half of the bridge
\begin{equation}
\label{bridge2} \aa \rightarrow K(\aa).
\end{equation}
The two halves of the bridge (\ref{bridge1}) and (\ref{bridge2})
fit well together since $K(C(X))$ is homeomorphic to $X$ while
$C(K(\aa))$ is isomorphic to $\aa$.

From a physics point of view the shift from space to algebra is
important. A point $x$ of $X$ can be seen as the object on which a
function $f$ is evaluated or equivalently, seen as characters,
points are objects to be evaluated on functions in order to give
numbers
\begin{equation}
\label{jolie}
 x(f) = f(x). \end{equation} The right-hand-side of (\ref{jolie}) refers
 to classical physics (first is space) while the
left-hand-side is closer to quantum mechanics (first are
observables). But it is not quantum mechanics since for the moment
we are dealing only with commutative obser\-vables.

Viewing points as a
characters-of-the-algebra-of-continuous-functions may sound a bit
complicated. But the algebraic point of view has the advantage to
be adaptable to the noncommutative framework. Namely starting from
a noncommutative algebra $\aa$, it is possible to build an object,
call it a non commutative space $Y$, such that $\aa$ plays the
role of functions on $Y$. Of course characters are not the
suitable tools to extract the geometric information from a
noncommutative algebra since characters precisely forget about the
noncommutativity. Instead one considers the {\it states} of the
algebra, that is to say the linear applications $\psi$ from $\aa$
to $\cc$ which are positive ($\psi(a^*a)\in\rr$) and of norm one
(which is equivalent to $\psi(\ii) = 1$ where $\ii$ is the unit of
$\aa$). The set $\ss(\aa)$ of states of a $C^*$-algebra with unit
is convex, which means that any state $\psi$ decomposes as
\begin{equation}
\psi = \lambda \omega_1 + (1-\lambda)\omega_2
\end{equation}
where $\omega_1$, $\omega_2$ are states and $\lambda\in [0,1]$.
The extremal points of $\ss(\aa)$, i.e. the states for which the
only convex combination is trivial ($\lambda =1$), are called {\it
pure states} of $\aa$. When the algebra is commutative characters
and pure states coincide. When the algebra is not commutative,
characters are not meaningful but pure states are available. That
is why from a topological point of view we consider the pure
states of $\aa$ as the "points" of the noncommutative space.

This is only topology and to do physics one needs much more than
topology. Especially dynamics requires a differential structure.
Is it possible to describe in an algebraic manner the differential
structure of a space ? The answer in general is no, the knowledge
of the spectrum of a differential operator is not enough to
recover the geometry of the underlying manifold("one cannot here
the shape of a drum"). Connes has shown that to recover geometry
from spectral datas, one needs more than the differential
structure, one has to consider also the spin structure.

\subsection{Spectral geometry}

A noncommutative geometry is given by a spectral triple
\begin{equation}
\aa, \hh, D
\end{equation}
where $\aa$ is an involutive algebra (commutative or not) with
unit, $\hh$ is an Hilbert space that carries a representation
$\pi$ of $\aa$ and $D$ is a selfadjoint operator acting on $\hh$,
generally unbounded. By definition these three elements compose a
spectral triple if and only if they satisfy a set of 7 properties
that pick out necessary and sufficient conditions allowing i. an
axiomatic definition of riemannian spin geometry in commutative
algebraic terms ii. an extension of this definition to the
noncommutative framework. Without entering the details that can be
found in the literature, let us simply indicate that the first
three conditions concern the analytical properties of $D$ and deal
with 1. the dimension of the space, 2. the smoothness of the
coordinates, 3. the "bundle nature" of a spin manifold. Then comes
4. a condition on the commutation of $D$ with the representation
that translates the fact that the Dirac operator is a first order
differential operator. Condition 5. is the algebraic formulation
of Poincare duality (duality between the $r^{\text{th}}$ and the
$(n\!-\!r)^{\text{th}}$ homology group of a $n$ dimensional
manifold). 6. concerns chirality and corresponds, in the
commutative case, to the orientation of the manifold. The last
condition 7. is the existence of a real structure, which allows
the lift of the frame bundle to the spin bundle.

One then checks that
\begin{equation}
\label{dirac} \aa=\cinf, \hh= L_2(M,S), D=
-i\gamma_\mu\partial_\mu
\end{equation}
(where $L_2(M,S)$ is the space of square integrable spinors over a
compact Riemannian spin manifold $M$ and $D$ is the usual Dirac
operator) satisfy the axioms of a spectral triple, hence a
Riemannian spin manifold is a noncommutative geometry. Conversely
a spectral triple with $\aa$ commutative fully determines a
Riemannian spin manifold (see Ref.\citelow{rennie,jgb} for a
detailed proof) whose geodesic distance is
\begin{equation}
\label{distance} d(x,y) = d(\omega_x, \omega_y) =
\underset{a\in\aa}{\sup}\left\{\abs{\omega_x(a) - \omega_y(a)}\; /
\norm{[D,\pi(a)]}\leq 1\right\}
\end{equation}
where $\omega_x$ is the point $x$ seen as a character of $\cinf$.
Thus Riemannian spin geometry is a particular case, commutative,
of an extended theory of geometry in spectral terms. In the
general case points are recovered as (pure) states $\omega,
\omega'$ of $\aa$ and formula (\ref{distance}) provides the metric
\begin{equation}
\label{distance2} d(\omega, \omega') =
\underset{a\in\aa}{\sup}\left\{\abs{\omega(a) - \omega'(a)}\; /
\norm{[D,\pi(a)]}\leq 1\right\}.
\end{equation}
Note that this formula is coherent for pure as well as non pure
states. For explicit computation, in the following we restrict to
pure states although one must be aware that, as shown by
Rieffel\cite{rieffel}, the knowledge of the distance function on
the state space is not determined in general by its restriction to
the set of pure states.

\subsection{Distance}

This is not difficult to check that the distance (\ref{distance})
associated to the geometry (\ref{dirac}) coincides with the
geodesic distance of the Riemannian structure of $M$. To
understand what is going on, one can look at the most basic
example by taking for $\aa$ the continuous functions on $\rr$
represented by multiplication on the space $\hh$ of square
integrable functions, and for $D$ the derivative with respect to
$x$. Then the operator $[D, \pi(f)]$ acts on $\psi\in\hh$ by
multiplication by the derivative $f'$ of $f$,
\begin{equation}
[D,\pi(f)]\psi = \frac{d}{dx} f\psi - f\frac{d}{dx}\psi =
(\frac{df}{dx})\psi
\end{equation}
hence
\begin{equation}
\norm{[D, \pi(f)]} = \underset{x\in\rr}{\sup}\abs{f'(x)}.
\end{equation}
The distance (\ref{distance}) writes
\begin{equation}
d(x,y) = d(\omega_x, \omega_y) = \underset{f\in
C(\rr)}{\sup}\left\{\abs{f(x) - f(y)}\; /
\underset{x\in\rr}{\sup}\abs{f'(x)}\leq 1\right\}
\end{equation}
which is nothing that $\abs{x-y}$, i.e. the geodesic distance in
$\rr$. The proof is identical for a manifold, except that the norm
of the commutator is given by the norm of the gradient of $f$.

Formula (\ref{distance2}) becomes interesting in situation where
the classical definition of the distance as the length of the
shortest path is no longer available. For instance noncommutative
geometry allows to describe a manifold whose disconnected
components are at finite distance from another. The simplest
example is given by the geometry
\begin{equation}
\label{2point}
\aa = \cc^2, \hh=\cc^2, D=\dm{cc} 0 & m \\
\overline{m} & 0 \fm
\end{equation}
where $m$ is a complex number and the representation of $(z_1,
z_2)\in\cc^2$ is diagonal
\begin{equation}
\pi(z_1, z_2) = \dm{cc} z_1 & 0\\ 0&z_2\fm .
\end{equation}
The two states of $\aa$ are
\begin{equation}
\omega_1 (z_1, z_2)\doteq z_1,\quad \omega_2 (z_1, z_2) \doteq z_2
\end{equation}
so we are dealing with a two point space. The computation of
(\ref{distance2}) is straightforward and yields
\begin{equation}
d(\omega_1, \omega_2)= \frac{1}{\abs{m}}.
\end{equation}

Next interesting example is $\aa=\m2$ represented over itself and
\begin{equation}
D\pi(a) = \Delta a + a\Delta\; \text{ with}\;  a\in\m2, \Delta \doteq\dm{cc} 0 & m \\
\overline{m} & 0 \fm.
\end{equation}
A pure state $\omega_\psi$ of $\m2$ is determined by a normalized
vector $\psi=\dm{c}\psi_1\\ \psi_2\fm $ of $\cc^2$,
\begin{equation}
\omega_\psi(a) = \scl{\psi}{a\psi}.
\end{equation}
By Hopf fibration any such vector is in one to one correspondence
with points of $S^2$
\begin{equation}
\psi \leftrightarrow \left\{\begin{array}{ccc}
x_\psi &=& \text{Re}(\overline{\psi_1}\psi_2)\\
y_\psi &=& \text{Im}(\overline{\psi_1}\psi_2)\\
z_\psi &=& \abs{\psi_1}^2 - \abs{\psi_2}^2\end{array}\right. .
\end{equation}
One then finds\cite{finite} that the distance $d$ between states
at different altitude (different value for $z$) is infinite while
it coincides with the euclidean distance of $S^2$ (up to a
constant factor $\frac{1}{\abs{m}}$) for states at the same
altitude.

It is also possible to describe spaces which are made of
continuous and discrete parts as, for instance, the product of the
spin geometry (\ref{dirac}) by the two point geometry
(\ref{2point}). Indeed given two spectral triples $(\aa_E, \hh_E,
D_E)$, $(\aa_I, \hh_I, D_I)$ the product
\begin{equation}
\label{product}
 \aa = \aa_E\ot\aa_I, \hh= \hh_E\ot\hh_I, D =
D_E\ot \ii_I + \Gamma_E\ot D_I \end{equation} ($\Gamma_E$ is the
chirality of the first geometry) is again a spectral triple. In
the case of a spin manifold $M$ multiplied by the two point
space(\ref{2point}), the space of pure states is a two sheet
models, two copies of $M$, and the distance coincides with the
geodesic distance of $M' = M\times [0,1]$ equipped with the metric
\begin{equation}
g_{ab} = \dm{cc} g_{\mu\nu} & 0 \\ 0 & \frac{1}{m^2}\fm
\end{equation}
where $g_{\mu\nu}$ is the metric on $M$. Note that although the
distance coincides with a geodesic distance, there is no geodesic
between two points on different sheets.

\section{Physics}

\subsection{Symmetry}

Trading spaces for algebras, one expects the symmetries of spaces
to have an algebraic translation. Inspired by the commutative case
in which
\begin{equation}
\text{Diff}(M) = \text{Aut}(\cinf),
\end{equation}
one will trade diffeomorphisms for automorphisms
\begin{equation} \label{cat}
\begin{array}{c}
\text{Diffeomorphisms of}\\
\text{the noncommutative space}\end{array} \Longleftrightarrow
\begin{array}{c}
\text{Automorphisms of}\\
\text{the algebra.}\end{array}
\end{equation}
An inner automorphism $\alpha_u$ is an automorphism implemented by
a unitary element ($u^* u = u u^* =\ii$) of $\aa$
\begin{equation}
\alpha_u(a) = u a u^* .
\end{equation}
Inner automorphisms form a group denoted $\text{In(\aa)}$. An
outer automorphism is a class in the quotient
\begin{equation}
\text{Out}(\aa) \doteq \text{Aut}(\aa) / \text{In}(\aa).
\end{equation}
When $\aa$ is commutative, $\text{In}(\aa)$ is trivial. Hence
inner automorphisms are a specificity of the noncommutative case
that remains hidden in the commutative case. They can be
interpreted as the noncommutative part of the "diffeomorphism
group" of the noncommutative space. In products of a manifold by a
finite dimensional geometry, like $\cinf\ot\m2$ in the precedent
section, the action of inner automorphism naturally yields a
scalar field which, as we will see later, can be identified as the
Higgs field of the standard model.

Inner automorphisms also makes the metric fluctuate. Given a
geometry $(\aa, \hh, D)$ with representation $\pi$ and real
structure $J$, one defines the action of $\text{In}(\aa)$ as
\begin{equation}
\pi \rightarrow \pi' = \pi\circ \alpha_u.
\end{equation}
This defines a new geometry $(\aa, \hh, D)$ where $\pi$ is
replaced by $\pi'$. A bit of algebra shows that this new geometry
is in fact equivalent to the geometry, with old representation
$\pi$, $(\aa, \hh, \dd)$ where
\begin{equation}
\dd= D + A + JAJ^{-1}
\end{equation} with
\begin{equation}
A \doteq u[D,u].
\end{equation}
$\dd$ is called the {\it covariant Dirac operator} because the
action of a unitary $u'$ on $(\aa, \hh, \dd)$ yields
\begin{equation}
\dd' = D + A' + JA'J^{-1}
\end{equation}
where
\begin{equation}
A' = u'Au'^* + u' [D,u'^*]
\end{equation}
which is similar to the transformation law of the potential in
gauge theory. The metric is said to fluctuate for, given two
states $\omega$, $\omega'$ of $\aa$, their distance $d$ in the
geometry $(\aa, \hh, D)$ does not equal in general their distance
$d_A$  in the geometry $(\aa, \hh, \dd)$ (there is no reason that
$\norm{[D,a]}$ equals $\norm{[\dd,a]}$). However the fluctuation
is covariant in the sense that
\begin{equation}
d_A(\omega, \omega') = d_{A'}(\omega\circ\alpha_{u'},
\omega'\circ\alpha_{u'}).
\end{equation}

Such fluctuations are a particular example of {\it inner
pertubations} given by
\begin{equation}
\dd = D + A + JAJ^{-1}
\end{equation}
where
\begin{equation}
\label{fluct}
 A = A^* = \underset{i}{\Sigma}\, a_i
[D,b_i]\quad\quad a_i, b_i \in\aa
\end{equation}
belongs to the set $\Omega_1$ of $1$-forms of the geometry.
Without entering the details (see Ref.\citelow{connes}) let us
just mention that the replacement of $D$ by $\dd$ corresponds to
the introduction of a connection. Trading the sections of a vector
bundle over $\mm$ for a finite projective $C(\mm)$-module (via
Serre-Swann theorem) one defines a connection on a $\aa$-module
$\ee$ as an application
\begin{equation}
\nabla: \ee \rightarrow \ee\ot \Omega_1
\end{equation}
satisfying the Leibniz rule as well as an hermitian condition (the
algebraic equi\-valent to the preservation of the metric by the
Levi-Civita connection). Taking for $\ee$ the algebra $\aa$
itself, the introduction of the connexion is then equivalent to
the replacement of $D$ by (\ref{fluct}).

\subsection{Standard model}

Taking the product of a manifold by a finite dimensional geometry
\begin{equation}
\aa_I = \cc \oplus \hhh \oplus M_3(\cc),\;  \hh_I = \cc^{90},\;
D_I
\end{equation}
where $90$ is the number of elementary fermions  ($6 \text{ quarks
} \times 3 \text{ colors } \times 2 \text{ chiralities }  + 3
\text{ leptons } \times 2 \text{ chiralities } + 3 \text{
neutrinos } = 45$ particles to which one adds $45$ antiparticles)
and $D_I$ is a matrix whose entries are the masses of the
elementary fermions and the coefficient of the unitary
Cabibbo-Kobayashi-Maskawa matrix, one can des\-cribe the geometry
of the standard model of elementary particles in spectral terms.
We refer to Ref.\citelow{schucker} for an updated presentation of
the subject. The inner fluctuations of the metric decompose in two
parts
\begin{equation}
A = -i \gamma^\mu \ot A_\mu - \gamma^5\ot H
\end{equation}
where $A_\mu$ is a vector field with value in the skew-adjoint
element of $\aa_I$ (i.e. in the Lie algebra of the unitaries of
$\aa_I$ which identifies with the gauge group of the standard
model) while $H$ is a scalar field with value in the internal
$1$-forms. Computing the asymptotic development of the spectral
action\cite{spectral}
\begin{equation}
\text{ Tr} (\theta (\frac{\dd^2}{\Lambda^2}))
\end{equation}
where $\theta$ is the characteristic function of the interval
$[0,1]$ and $\Lambda$ a cut-off , one finds the Einstein-Hilbert
action (with euclidean signature) + a Weyl term + the bosonic part
of the standard model. The Higgs field identifies to $H$ and get a
geometrical meaning in terms of the scalar part of the fluctuation
of the metric. The metric interpretation of the Higgs fields has
been fully elucidated in Ref.\citelow{kk}. Pure states $\aa$
defines a two sheet model, two copies of $M$, one indexed by the
pure state of $\cc$, the other one by the pure state of $\hhh$
(algebra of quaternions). The pure states of $M_3(\cc)$ are
degenerated from a metric point of view. Writing $x_\cc$, $y_\hhh$
two points on different sheets, it turns out that for a pure
scalar fluctuation ($A_\mu = 0$) the noncommutative distance
(\ref{distance2}) coincide with the geodesic distance of $M\times
[0,1]$ with metric
\begin{equation}
g^{ab} = \dm{cc} g^{\mu\nu} & 0 \\ 0 & g^{tt}(x)\fm
\end{equation}
where the extra metric component $g^{tt}$ is given by the Higgs
doublet $\dm{c} 1 + h_1\\ h_2 \fm$ and the norm of the mass
matrix, i.e. the mass of the quark top,
\begin{equation}
g^{tt}(x) = ( \abs{1 + h_1(x)}^2 + \abs{h_2(x)}^2) m_{top}^2.
\end{equation}
Once again let us emphasize that there is nothing between the two
sheets (so no geodesic between points on different sheets). The
distance is finite although the internal space is discrete.

\subsection{Thermal time hypothesis}

We have seen in the precedent subsection that inner perturbations
yield the bosonic part of the standard model. Outer automorphisms
also have a nice physical interpretation in terms of dynamics. The
starting point is the observation that a Von Neumann algebra $\aa$
(acting on a Hilbert space $\hh$) is a dynamical object, in the
sense that it comes equipped with a canonical one-parameter group
of outer automorphism, the modular group of Tomita-Takesaki
\begin{eqnarray}
s&\mapsto&\sigma^s\in\text{Aut}(\aa)\\
& &\sigma^s(a) = \Delta^{is} a \Delta^{-is}.
\end{eqnarray}
$\Delta$ is given by the polar decomposition of the closure of the
operator $S$
\begin{equation}
Sa\Omega = a^* \Omega
\end{equation}
where $\Omega$ is a vector cyclic and separating for the action of
$\aa$. $S$, hence $\Delta$ hence $\sigma$ depends on the initial
choice of $\Omega$. The remarkable point (co-cycle Radon-Nikodym
theorem\cite{radon}) is that $\sigma^s$ depends on $\Omega$ only
modulo inner automorphisms. Hence there is a unique one parameter
group of outer automorphisms associated to $\aa$ via the modular
theory. Let us fix one representant $\sigma$ in this unique class
of equivalence, and write $\Omega$ the corresponding vector. Then
$\sigma$ has the remarkable property that it satisfies with
respect to $\Omega$ the same properties as the time evolution
$\alpha$ with respect to a thermal equilibrium state $\omega$ at
inverse temperature $\beta$, namely the KMS condition\cite{haag}
\begin{equation}
\label{kms}
 \omega(A\alpha^t(B)) = \omega(\alpha^{t+i\beta}(B)A).
 \end{equation}
Here $A$, $B$ are observables of a thermodynamical system with
Hamiltonian $H$, $\omega$ is a Gibbs state $\omega(A) = Z^{-1}
\text{Tr} (A e^{-\beta H})$ with $Z$ the partition function and
$\alpha^t(A)= e^{-iHt}Ae^{iHt}$. In fact one has
\begin{equation}
 \scl{\Omega}{a\sigma^s b\Omega}= \scl{\Omega}{(\alpha^{s -i}b)a)}
 \end{equation}
 which yields the KMS condition (\ref{kms}) if we put
 \begin{equation}
\sigma^s = \alpha^{-\beta t}.
 \end{equation}
 Hence an equilibrium state at inverse temperature $\beta$ is a state such that its modular
 group $\sigma^s$ coincides with the time flow $\alpha^t$, the parameter $s$ being related to the time
 $t$ by
 \begin{equation}
 s = -\beta t.
 \end{equation}

The modular group is a formal time evolution for the state defined
by the vector $\Omega$. Connes and Rovelli\cite{tth} have
suggested that this evolution may have a physical meaning. The
{\it thermal time hypothesis} demands that the modular flow
determined by the statistical state of a real physical system
coincides with what we perceive as the physical flow of time. This
hypothesis was initially motivated by the problem of time in
quantum gravity\cite{ish}. For the time being it has been tested
in semi-classical situations where a geometrical background
already provides an independent notion of temporal flow. In this
case the hypothesis demands that the ratio of the rates of the two
flows (geometrical and modular) be identified as the temperature
of the state. The Unruh effect is an example in which the thermal
time hypothesis is realized. Let us recall  that Unruh effect is
the theoretical observation that the vacuum state $\Omega$ of a
quantum field theory on Minkovski spacetime looks like a thermal
equilibrium state for an uniformly accelerated observer $\oo$ with
acceleration $a$. The observed temperature is the Davis-Unruh
temperature $T_{U}=\hbar a/2\pi k_{b}c$. Among the many
derivations of Unruh effect\cite{sewell} one is based on the
geometrical properties of the region causally connected to the
world line of $\oo$, namely the Rindler wedge $W$ ($\abs{t}< {\bf
\norm{x}}$). The modular group defined by $\Omega$ on the algebra
of local observables on $W$ has a geometrical
action\cite{bisognano} which coincides with the proper time flow
of $\oo$. The proportionality constant between the two flows is
$T_U$ and is interpreted as the temperature of the vacuum seen by
$\oo$. Note that a similar analysis has been developed in
Ref.\citelow{diamonds} for the modular flow associated to the
causal horizon of a non eternal uniformly accelerated observer,
namely a diamond region $D_L$ ($\abs{t} + \norm{{\bf x}} < L$ with
$L$ a constant).

Here a time flow is given (the proper time flow of $\oo$) as well
as a state (the vacuum $\Omega$) and the coincidence between the
time flow and the modular flow of $W$ yields the interpretation of
the ratio as a temperature
$$\left\{ \begin{array}{l} \text{time}\\ \text{state}\end{array}\right.
\longrightarrow \text{temperature}.
$$
The thermal time hypothesis makes the opposite analysis: assuming
the vacuum is thermal with temperature $T_U$, then physical time
is given by the modular flow and it turns out that this time
coincides with the proper time of $\oo$
$$\left\{ \begin{array}{l} \text{state}\\ \text{temperature}\end{array}\right.
\longrightarrow \text{time}.
$$
This shift in the point of view makes the thermal time hypothesis
a interesting tool in quantum gravity for it may allow to extract
from a fully covariant quantum formulation of gravity our
(strongly non covariant!) intuition of flow of time. Indeed
assuming that covariance of general relativity is preserved at the
quantum level, then one has the freedom to pick out from the
surrounding superposition of states of the gravitational field any
particular direction as {\it the} time direction. The thermal time
hypothesis gives a way to make this freedom compatible with our
local intuition of physical flow of time, by making time a state
dependant notion. For the moment the hypothesis has not been
applied on this context, because we are still lacking of a
definite algebra of quantum gravity observables to begin the
analysis (i.e. try to compute the modular flow and possibly
interpret it in dynamical terms). The situation may change since
there are now some candidates as algebras of observables in loop
quantum gravity.

\section{Quantum gravity ?}

One generally assumes that quantum gravity should yield a non
continuous structure of spacetime at Planck scale. Recently it has
been underlined that such a structure may not be out of the reach
of experimental measurement\cite{giac}, first because it induces a
modification of the usual relativistic dispersion relation which
may have a significant effect on the propagation of high energy
cosmic rays, second because the "fuzzyness" of space may yield a
characteristic source of noise in gravitational waves experiment.
However discretness of spacetime might not be only a quantum
gravity effect since noncommutative geometry provides the
spacetime of the standard model with a discrete structure already
at the classical level (i.e. classical gravitation). Therefore it
could be interesting to adapt to the two sheet model of the
standard model the analysis of quantum gravity phenomenology
developed for quantum spaces. For instance one could study the
propagation of some signal on the two sheet model, or export the
distance formula in quantum spaces so that to make a quantitative
analysis of the "fuzzyness" of spacetime.

The fact that discreteness of spacetime is not necessarily a
quantum gravitational effect makes the link quantum
gravity/quantum spaces/noncommutative geometry still more
complicated. It is likely that quantum gravity will have to do
with a noncommutative structure of spacetime. It is certainly to
soon to know whether it should be via the discreteness of
spacetime, via the richness of the "diffeomorphism group" of a
noncommutative space, or via something else.


\end{document}